\documentclass[cits]{PoS}

\usepackage{amssymb, fontenc, times, mathptmx, graphicx,amsmath,amsfonts,slashed}

\newcommand{\beq}{\begin{equation}}
\newcommand{\eeq}{\end{equation}}
\newcommand{\Order}{\mathcal{O}}

\newcommand{\mpp}{m_{\rm p}}
\newcommand{\mn}{m_{\rm n}}
\newcommand{\mpi}{M_{\pi}}
\newcommand{\mpii}{M_{\pi^0}}

\newcommand{\Fpi}{F_\pi}

\title{Isospin-breaking corrections to the pion--nucleon scattering lengths}

\ShortTitle{Isospin-breaking corrections to the pion--nucleon scattering lengths}

\author{\speaker{Martin Hoferichter}%
         \\
        Helmholtz--Institut f\"ur Strahlen- und Kernphysik (Theorie) 
   and Bethe Center for Theoretical Physics, Universit\"at Bonn, D-53115 Bonn, Germany\\
        E-mail: \email{hoferichter@hiskp.uni-bonn.de}}

\author{Bastian Kubis%
         \\
        Helmholtz--Institut f\"ur Strahlen- und Kernphysik (Theorie) 
   and Bethe Center for Theoretical Physics, Universit\"at Bonn, D-53115 Bonn, Germany\\
        E-mail: \email{kubis@hiskp.uni-bonn.de}}       
        
\author{Ulf-G. Mei{\ss}ner%
         \\
        Helmholtz--Institut f\"ur Strahlen- und Kernphysik (Theorie) 
   and Bethe Center for Theoretical Physics, Universit\"at Bonn, D-53115 Bonn, Germany\\
   and\\
   Institut f\"ur Kernphysik (Theorie), Institute for Advanced Simulation, 
   and J\"ulich Center for Hadron Physics, Forschungszentrum J\"ulich, D-52425  J\"ulich, Germany\\
        E-mail: \email{meissner@hiskp.uni-bonn.de}}

\abstract{We analyze isospin breaking through quark mass differences and virtual photons 
in the pion--nucleon scattering lengths in all physical channels in the
framework of covariant baryon chiral perturbation theory. The so-called triangle relation 
is found to be violated by about 1.5 \%. We encounter a substantial isospin-breaking 
correction to neutral-pion--nucleon scattering beyond Weinberg's prediction due to a cusp effect. 
Finally, the application to hadronic atoms is briefly discussed.}

\FullConference{6th International Workshop on Chiral Dynamics\\
		 July 6-10 2009\\
		 Bern, Switzerland}

\begin{document}

\section{Introduction}

Isospin violation in the Standard Model is driven by strong and 
electromagnetic interactions, that is by the differences in the
light quark masses and charges, respectively. Already in~\cite{Weinberg77} Weinberg stressed
that the pion--nucleon scattering lengths offer a particularly
good testing ground for strong isospin violation, predicting e.g.\ a large effect in the difference of the neutral-pion--nucleon scattering lengths $a_{\pi^0 p}-a_{\pi^0 n}$. Isospin violation in $\pi N$ scattering
was addressed in the framework of heavy-baryon chiral perturbation
theory (ChPT) in a series of papers about a decade ago 
\cite{MS97,FMS99,MM99,FM01b,FM01}.
Recently, new interest arose in high-precision calculations of the pion--nucleon 
scattering lengths. First, corrections for isospin violation are an essential ingredient to extract 
the isoscalar and isovector  $\pi N$ scattering lengths $a^+$ and $a^-$
from pionic hydrogen ($\pi H$) and deuterium ($\pi D$) measurements  to high precision.
The isospin-breaking corrections to $a_{\pi^- p\to \pi^- p}$ needed for the ground state of $\pi H$ were  calculated in~\cite{GR02} at third chiral order. A consistent description requires
the knowledge of $a_{\pi^- p\to \pi^0 n}$ (width of $\pi H$) and $a_{\pi^- n\to \pi^- n}$ (ground state of $\pi D$) at the same accuracy.
In the analysis of $\pi D$ isospin violation is particularly
important, since the $\pi d$ scattering length at leading order is
proportional to the small isoscalar scattering length $a^+$ 
and therefore chirally suppressed ~\cite{MRR06}. Second, threshold pion photoproduction offers
the unique possibility of measuring the so far undetermined $\pi^0 p$ scattering
length and gives access to the charge exchange scattering length $a_{\pi^+ n
\to\pi^0 p}$ \cite{Bernstein:1998ip,Bernstein:2009dc}. Such measurements are becoming 
feasible at HI$\gamma$S and at MAMI.
In view of these developments, we have extended
the work of~\cite{GR02} to {\it all} charge channels in
pion--nucleon scattering in~\cite{HKM09}.

\section{Formalism}

The calculation of the scattering lengths is performed at $\Order(p^3)$ in manifestly covariant baryon ChPT \cite{BL99} and at first order in the isospin-breaking parameter $\delta=\Order(e^2,m_{\rm d}-m_{\rm u})$, using the effective Lagrangian for nucleons, pions, and virtual photons, as constructed in~\cite{GR02}. We denote the masses of proton, neutron, charged and neutral pion by $\mpp$, $\mn$, $\mpi$, and $\mpii$, respectively, and define the isospin limit by the charged particles. The mass differences are expressed by $\Delta_\pi=\mpi^2-\mpii^2$ and $\Delta_{\rm N}=\mn-\mpp$.

\begin{figure}
\begin{center}
\includegraphics[width=0.7\linewidth,clip]{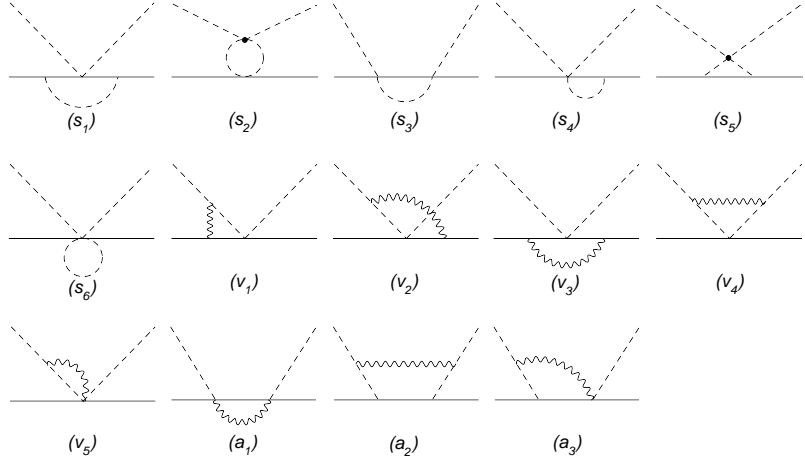}
\end{center}
\caption{One-loop topologies for $\pi N$ scattering at threshold. 
Solid, dashed, and wiggly lines, denote nucleons, pions, and photons, respectively.
Crossed diagrams and diagrams contributing via wave function renormalization only are not shown.}
\label{fig:diagrams}
\end{figure}

The one-loop topologies are depicted in Fig.~\ref{fig:diagrams}.
As soon as we take into account virtual photons, 
we have to specify how to deal with the threshold divergences:
first of all, we subtract all one-photon-reducible diagrams, since they diverge $\sim 1/t$ ($s$ and $t$ are the usual Mandelstam variables), 
and denote the resulting amplitude by $\tilde{T}$. 
The additional divergences due to photon loops may be regularized in the form
\beq
e^{iQ\alpha\theta_{\rm C}(|\mathbf{p}|)}\tilde{T}\Big|_{|\mathbf{p}|\rightarrow 0}=\frac{\beta_1}{|\mathbf{p}|}+\beta_2\log\frac{|\mathbf{p}|}{\mu_{\rm c}}+T_{\rm thr}+\Order(|\mathbf{p}|),
\eeq
where $\mathbf{p}$ denotes the center-of-mass momentum, $\alpha={e^2}/{4\pi}$ the fine structure constant,
$\theta_{\rm C}(|\mathbf{p}|)$ the infrared divergent Coulomb phase given by
\beq
\theta_{\rm C}(|\mathbf{p}|)=-\frac{\mu_{\rm c}}{|\mathbf{p}|}\log\frac{m_\gamma}{2|\mathbf{p}|} \,,
\eeq
$\mu_{\rm c}={\mpp\mpi}/{(\mpp+\mpi)}$ the reduced mass of the incoming particles, 
and $Q$ accounts for the charges of the particles involved. The term $\propto 1/|\mathbf{p}|$ referred to as Coulomb pole is solely generated by $(v_1)$ at this order, while $\beta_2$ only enters at two-loop level. The scattering lengths are finally given by
\beq
a=\frac{T_{\rm thr}}{8\pi\sqrt{s}}.
\eeq
It turns out that for the charged-pion elastic channels the triangle graph $(s_5)$ yields a very large contribution. In addition, $(s_3)$ can generate cusp effects, which are proportional to $\sqrt{\delta}$ and thus enhanced by $\sqrt{\delta}$.
For the analytical results and more details of the calculation we refer to \cite{HKM09}.

\section{Numerical results}

In the isospin limit, the $\pi N$ scattering lengths are solely determined by $a^+$ and $a^-$. Subtracting these contributions, we obtain the following isospin-breaking shifts (in units of $10^{-3}\mpi^{-1}$):

\vspace{3mm}

\begin{tabular}{c|c|c|c|c}
 		isospin limit & channel & shift & channel & shift \\\hline
 		 $a^++a^-$ & $\pi^-p\rightarrow \pi^-p$ & $-3.4^{+4.3}_{-6.5}+5.0 i$ & $\pi^+n\rightarrow \pi^+n$ & $-4.3^{+4.3}_{-6.5}+6.0 i$ \\
 		 $a^+-a^-$ & $\pi^+p\rightarrow \pi^+p$ & $-5.3^{+4.3}_{-6.5}$ & 	$\pi^-n\rightarrow \pi^-n$  & $-6.2^{+4.3}_{-6.5}$ \\
 		 $-\sqrt{2}\,a^-$ & $\pi^-p\rightarrow \pi^0n$ & $0.4\pm 0.9$ &  	$\pi^+n\rightarrow \pi^0 p$ & $2.3\pm 0.9$\\
 		 $a^+$ & $\pi^0p\rightarrow \pi^0p$ & $-5.2\pm 0.2$ & 		$\pi^0n\rightarrow \pi^0 n$ &  $-1.8\pm 0.2$
\end{tabular} 

\vspace{3mm}

The precise values of the low-energy constants as well as a detailed estimate of the theoretical uncertainties can be found in \cite{HKM09}. Our result for the triangle relation, which vanishes in the isospin limit and is thus a convenient way to quantify isospin violation in terms of measurable quantities, reads
\beq
R=2\frac{ a_{\pi^+ p\rightarrow \pi^+ p}- a_{\pi^- p\rightarrow \pi^- p}-\sqrt{2}\, a_{\pi^- p\rightarrow \pi^0 n}}{ a_{\pi^+ p\rightarrow \pi^+ p}- a_{\pi^- p\rightarrow \pi^- p}+\sqrt{2}\, a_{\pi^- p\rightarrow \pi^0 n}}=(1.5\pm 1.1)\,\%.
\eeq
The difference between the elastic neutral-pion--nucleon scattering lengths is found to be
\begin{align}
a_{\pi^0p}-a_{\pi^0n}&=\frac{\mpp}{4\pi(\mpp+\mpi)} \biggl\{\frac{4c_5 B(m_{\rm d}-m_{\rm u})}{\Fpi^2}-\frac{\mpi^2}{8\pi\Fpi^4}\Big(\sqrt{\Delta_\pi+2\mpi\Delta_{\rm N}}-\sqrt{\Delta_\pi-2\mpi\Delta_{\rm N}}\Big)\biggr\}\notag\\
	&=\big((-2.3\pm 0.4)-1.1\big)\cdot 10^{-3}\mpi^{-1}=(-3.4\pm 0.4)\cdot 10^{-3}\mpi^{-1}.
\end{align} 
The first term was already given by Weinberg in \cite{Weinberg77}, while the second one is due to a cusp effect and contributes roughly one third to the final result.

\section{Application to hadronic atoms}

\begin{figure}
\begin{center}
\includegraphics[width=0.7\linewidth,clip]{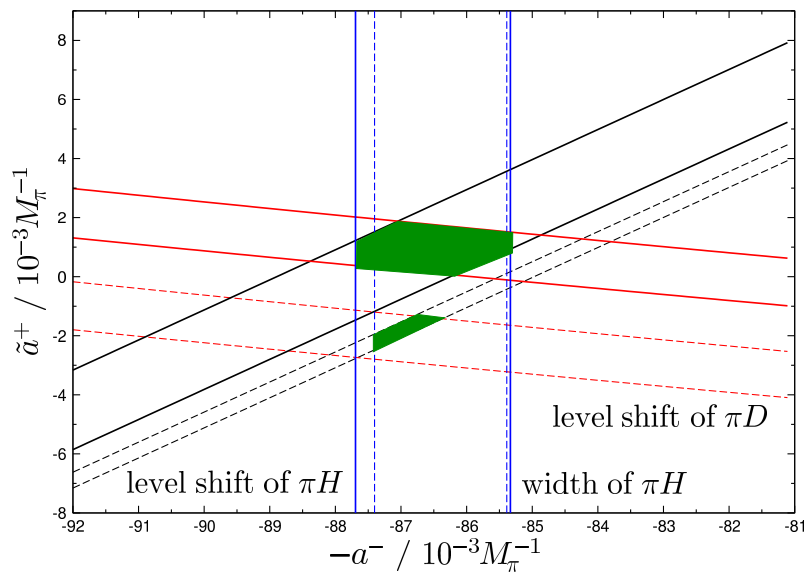}
\end{center}
\caption{Constraints on $a^-$ and $\tilde{a}^+$ provided by the level shift of $\pi H$ and $\pi D$ and the width of $\pi H$. Solid (dashed) bands refer to isospin-breaking corrections at $\Order(p^3)$ ($\Order(p^2)$), respectively.}
\label{fig:bands}
\end{figure}

$a^+$ and $a^-$ can be related to the strong shift of the ground state of $\pi H$ and $\pi D$ and to the width of $\pi H$ via Deser-type formulae \cite{LR00, Zemp,MRR05,GLR07},  to which isospin-breaking corrections are an essential ingredient. The impact of our isospin-breaking corrections on the extraction of $a^-$ and 
\beq
\tilde{a}^+ =a^++ \frac{\mpp}{4\pi(\mpp+\mpi)}
\bigg\{\frac{4\Delta_\pi}{\Fpi^2}c_1-2e^2f_1 \bigg\}
\eeq
is displayed in Fig.~\ref{fig:bands}:\footnote{Experimental input: level shift of $\pi H$: $\epsilon_{1s}=(-7.120\pm 0.017)\,{\rm eV}$ \cite{Gotta05}, width of $\pi H$: $\Gamma_{1s}=(0.823\pm 0.019)\,{\rm eV}$ \cite{Gotta08}, and $\pi^- d$ scattering length: $a_{\pi d}=(-0.0261\pm 0.0005)\,\mpi^{-1}$\cite{Hauser98}.} the constraint due the width of $\pi H$ barely changes, while the the bands for the level shift of $\pi H$ and $\pi D$ significantly move upwards when going from $\Order(p^2)$ to $\Order(p^3)$, corresponding to the small corrections to the charge exchange reaction and a large shift in the charged-pion elastic channels due to the triangle graph alluded to above. 

However, Fig.~\ref{fig:bands} does not provide a complete picture at $\Order(p^3)$: the few-body contributions to $a_{\pi d}$ 
\cite{BBEMP02, Baru1, Baru2} are still based on the assumption of isospin symmetry. It remains to be seen whether the nice consistency between the three bands persists once isospin violation in the few-body part is included.

\section{Summary and outlook}

We have systematically analyzed isospin violation in the $\pi N$ scattering
lengths in all channels, including an estimate of the
theoretical uncertainties. We find that isospin breaking is quite small in $\pi^- p\rightarrow \pi^0 n$, 
at the order of one percent at most, whereas the
charged-pion elastic channels display more sizeable effects on the few-percent level. 
In particular, the triangle relation 
is violated by about 1.5\% consistent with earlier findings in heavy-baryon
ChPT and inconsistent with the 5--7\% deviation extracted from the data at
lowest pion momenta in \cite{Gibbs:1995dm,Matsinos:1997pb}.
In addition, we find a substantial isospin-breaking correction to the neutral-pion--proton scattering
length. Finally, we have shown that in a full $\Order(p^3)$ calculation the value of the isoscalar scattering length will increase as compared to previous analyses of hadronic atoms \cite{MRR06}.

An extension of this analysis beyond threshold for  $\pi^\pm p\rightarrow \pi^\pm p$ and $\pi^- p\rightarrow \pi^0 n$ can be found in~\cite{HKM09_above}, while a full calculation of isospin violation in the few-body contributions to the $\pi d$ scattering length will be addressed in \cite{BHHKNP09}.

\acknowledgments

Partial financial support by the Helmholtz Association through funds provided
to the virtual institute ``Spin and strong QCD'' (VH-VI-231),
by the European Community-Research Infrastructure Integrating Activity 
``Study of Strongly Interacting Matter''
(acronym HadronPhysics2, Grant Agreement n.~227431) under the Seventh 
Framework Programme of the EU,
by DFG (SFB/TR 16, ``Subnuclear Structure of Matter'') and by the Bonn-Cologne Graduate School of Physics and Astronomy is gratefully
acknowledged.


\begin{thebibliography}{99}
\bibitem{Weinberg77}
  S.~Weinberg,
  \emph{The Problem Of Mass},
  \emph{Trans.\ New York Acad.\ Sci.\ }  {\bf 38} (1977) 185.

\bibitem{MS97}
  U.-G.~Mei{\ss}ner and S.~Steininger,
  \emph{Isospin violation in pion-nucleon scattering},
  \emph{Phys.\ Lett.\  B} {\bf 419} (1998) 403
  [{\tt hep-ph/9709453}].

\bibitem{FMS99}
  N.~Fettes, U.-G.~Mei{\ss}ner and S.~Steininger,
  \emph{On the size of isospin violation in low-energy pion nucleon scattering},
  \emph{Phys.\ Lett.\  B} {\bf 451} (1999) 233
  [{\tt hep-ph/9811366}].

\bibitem{MM99}
  G.~M\"uller and U.-G.~Mei{\ss}ner,
  \emph{Virtual photons in baryon chiral perturbation theory},
  \emph{Nucl.\ Phys.\  B} {\bf 556} (1999) 265
  [{\tt hep-ph/9903375}].

\bibitem{FM01b}
  N.~Fettes and U.-G.~Mei{\ss}ner,
  \emph{Towards an understanding of isospin violation in pion nucleon
  scattering},
  \emph{Phys.\ Rev.\  C} {\bf 63} (2001) 045201
  [{\tt hep-ph/0008181}].
    
\bibitem{FM01}
  N.~Fettes and U.-G.~Mei{\ss}ner,
  \emph{Complete analysis of pion nucleon scattering in chiral perturbation  theory
  to third order},
  \emph{Nucl.\ Phys.\  A} {\bf 693} (2001) 693
  [{\tt hep-ph/0101030}].

\bibitem{GR02}
  J.~Gasser, M.~A.~Ivanov, E.~Lipartia, M.~Moj\v{z}i\v{s} and A.~Rusetsky,
  \emph{Ground-state energy of pionic hydrogen to one loop},
  \emph{Eur.\ Phys.\ J.\  C} {\bf 26} (2002) 13
  [{\tt hep-ph/0206068}].
  
\bibitem{MRR06}
  U.-G.~Mei{\ss}ner, U.~Raha and A.~Rusetsky,
  \emph{Isospin-breaking corrections in the pion deuteron scattering length},
  \emph{Phys.\ Lett.\   B} {\bf 639} (2006) 478
  [{\tt nucl-th/0512035}].

\bibitem{Bernstein:1998ip}
  A.~M.~Bernstein,
  \emph{Light quark mass difference and isospin breaking in electromagnetic  pion
  production},
  \emph{Phys.\ Lett.\  B} {\bf 442} (1998) 20
  [{\tt hep-ph/9810376}].

\bibitem{Bernstein:2009dc}
  A.~M.~Bernstein, M.~W.~Ahmed, S.~Stave, Y.~K.~Wu and H.~R.~Weller,
  \emph{Chiral Dynamics in Photo-Pion Physics: Theory, Experiment, and Future
  Studies at the HI$\gamma$S Facility},
  {\tt nucl-ex/0902.3650}.
  
\bibitem{HKM09}
  M.~Hoferichter, B.~Kubis and U.-G.~Mei{\ss}ner,
  \emph{Isospin breaking in the pion--nucleon scattering lengths},
  \emph{Phys.\ Lett.\  B} {\bf 678} (2009) 65
  [{\tt 0903.3890 [hep-ph]}].


  


\bibitem{BL99}
  T.~Becher and H.~Leutwyler,
 \emph{Baryon chiral perturbation theory in manifestly Lorentz invariant form},
  \emph{Eur.\ Phys.\ J.\  C} {\bf 9} (1999) 643
  [{\tt hep-ph/9901384}].


\bibitem{LR00}
  V.~E.~Lyubovitskij and A.~Rusetsky,
  \emph{$\pi^- p$ atom in ChPT: Strong energy-level shift},
  \emph{Phys.\ Lett.\  B} {\bf 494} (2000) 9
  [{\tt hep-ph/0009206}].
  
\bibitem{Zemp}
P.~Zemp, \emph{Pionic hydrogen in QCD+QED: decay width at NNLO}, PhD thesis, University of Bern, 2004.  
  
\bibitem{MRR05}
  U.-G.~Mei{\ss}ner, U.~Raha and A.~Rusetsky,
  \emph{The pion nucleon scattering lengths from pionic deuterium},
  \emph{Eur.\ Phys.\ J.\  C} {\bf 41} (2005) 213
  [{\tt nucl-th/0501073}].

\bibitem{GLR07}
  J.~Gasser, V.~E.~Lyubovitskij and A.~Rusetsky,
  \emph{Hadronic atoms in QCD + QED},
  \emph{Phys.\ Rept.\ }  {\bf 456} (2008) 167
  [{\tt 0711.3522 [hep-ph]}].
  

\bibitem{Gotta05}
  D.~Gotta  [Pionic Hydrogen Collaboration],
  \emph{Pionic hydrogen},
  \emph{Int.\ J.\ Mod.\ Phys.\  A} {\bf 20} (2005) 349.

\bibitem{Gotta08}
  D.~Gotta {\it et al.},
  \emph{Pionic hydrogen},
  \emph{Lect.\ Notes Phys.\ } {\bf 745} (2008) 165.


\bibitem{Hauser98}
  P.~Hauser {\it et al.},
  \emph{New precision measurement of the pionic deuterium s-wave strong
  interaction parameters},
  \emph{Phys.\ Rev.\  C} {\bf 58} (1998) 1869.
 
 
\bibitem{BBEMP02}
  S.~R.~Beane, V.~Bernard, E.~Epelbaum, U.-G.~Mei{\ss}ner and D.~R.~Phillips,
  \emph{The S-wave pion nucleon scattering lengths from pionic atoms using
  effective field theory},
  \emph{Nucl.\ Phys.\  A} {\bf 720} (2003) 399
  [{\tt hep-ph/0206219}].

\bibitem{Baru1}
  V.~Lensky, V.~Baru, J.~Haidenbauer, C.~Hanhart, A.~E.~Kudryavtsev and U.-G.~Mei{\ss}ner,
  \emph{Dispersive and absorptive corrections to the pion deuteron scattering
  length},
  \emph{Phys.\ Lett.\  B} {\bf 648} (2007) 46
  [{\tt nucl-th/0608042}].


\bibitem{Baru2}
  V.~Baru, J.~Haidenbauer, C.~Hanhart, A.~E.~Kudryavtsev, V.~Lensky and U.-G.~Mei{\ss}ner,
  \emph{Role of the Delta(1232) in pion-deuteron scattering at threshold within
  chiral effective field theory},
  \emph{Phys.\ Lett.\  B} {\bf 659} (2008) 184
  [{\tt 0706.4023 [nucl-th]}].




\bibitem{Gibbs:1995dm}
  W.~R.~Gibbs, L.~Ai and W.~B.~Kaufmann,
  \emph{Isospin Breaking In Low-Energy Pion Nucleon Scattering},
  \emph{Phys.\ Rev.\ Lett.\ }  {\bf 74} (1995) 3740.

\bibitem{Matsinos:1997pb}
  E.~Matsinos,
  \emph{Isospin violation in the pi N system at low energies},
  \emph{Phys.\ Rev.\  C} {\bf 56} (1997) 3014.
  
\bibitem{HKM09_above}
 M.~Hoferichter, B.~Kubis and U.-G.~Mei{\ss}ner,
  \emph{Isospin violation in low-energy  pion--nucleon scattering revisited}, {\tt 0909.4390 [hep-ph]}.
  
\bibitem{BHHKNP09}
  V.~Baru, C.~Hanhart, M.~Hoferichter, B.~Kubis, A.~Nogga, D.~R.~Phillips, in preparation.  

\end{thebibliography}
\end{document}